\RequirePackage{fix-cm}
\documentclass[twocolumn,epjc3]{svjour3}  
\smartqed  
\RequirePackage{graphicx}
\RequirePackage{comment}
\usepackage{color}
\usepackage{lineno}
\usepackage{url}
\usepackage{amsmath}

\newcommand{\DBD}{0$\nu$DBD}

\newcommand{\Se}{$^{82}\mathrm{Se}$}
\newcommand{\THO}{$^{232}\mathrm{Th}$}
\newcommand{\qvalue}{2997.9$\pm$0.3\,keV}
\newcommand{\duenu}{T$_{1/2}^{2\nu}$=  [9.39 $\pm$ 0.17 (stat) $\pm$ 0.58 (syst)] $\times$ 10$^{19}$\,yr  }

\newcommand{\CupidZ}{CUPID-0}
\newcommand{\EnrZnSe}{Zn$^{82}$Se}
\newcommand{\dataset}{DataSet}

\newcommand{\exposure}{5.74\,kg$\cdot$yr}
\newcommand{\exposureSe}{2.24$\times$10$^{25}$\,emitters$\cdot$yr}

\newcommand{\LimitFirst}            {$\Gamma$($^{82}$Se $\rightarrow ^{82}$Kr$_{0_1^+}$)$<$8.55$\times$10$^{-24}$\,yr$^{-1}$}

\newcommand{\LimitSecondTwo}{$\Gamma$($^{82}$Se $\rightarrow ^{82}$Kr$_{2_2^+}$)$<$8.25$\times$10$^{-24}$\,yr$^{-1}$}

\newcommand{\LimitFirstGe}{$\Gamma$($^{82}$Se $\rightarrow ^{82}$Kr$_{0_1^+}$)$<$2.0$\times$10$^{-23}$\,yr$^{-1}$}
\newcommand{\LimitSecondOneGe}{$\Gamma$($^{82}$Se $\rightarrow ^{82}$Kr$_{2_1^+}$)$<$5.3$\times$10$^{-23}$\,yr$^{-1}$}
\newcommand{\LimitSecondTwoGe}{$\Gamma$($^{82}$Se $\rightarrow ^{82}$Kr$_{2_2^+}$)$<$6.7$\times$10$^{-23}$\,yr$^{-1}$}

\journalname{Eur. Phys. J. C}

\begin{document}

\title{Search of the neutrino-less double beta decay of $^{82}$Se into the excited states of $^{82}$Kr with CUPID-0}
\author{O.~Azzolini\thanksref{Legnaro}
\and{M.~T.~Barrera\thanksref{Legnaro}
\and J.~W.~Beeman\thanksref{LBNL}
\and F.~Bellini\thanksref{Roma,INFNRoma}
\and M.~Beretta\thanksref{MIB,INFNMiB}
\and M.~Biassoni\thanksref{INFNMiB}
\and E.~Bossio\thanksref{Roma,INFNRoma} 
\and C.~Brofferio\thanksref{MIB,INFNMiB}
\and C.~Bucci\thanksref{LNGS}
\and L.~Canonica\thanksref{LNGS,e2} 
\and S.~Capelli\thanksref{MIB,INFNMiB}
\and L.~Cardani\thanksref{INFNRoma,e1}
\and P.~Carniti\thanksref{MIB,INFNMiB}
\and N.~Casali\thanksref{INFNRoma}
\and L.~Cassina\thanksref{MIB,INFNMiB}
\and M.~Clemenza\thanksref{MIB,INFNMiB}
\and O.~Cremonesi\thanksref{INFNMiB}
\and A.~Cruciani\thanksref{INFNRoma}
\and A.~D'Addabbo\thanksref{LNGS,GSSI}
\and I.~Dafinei\thanksref{INFNRoma}
\and S.~Di~Domizio\thanksref{Genova,INFNGenova}
\and F.~Ferroni\thanksref{Roma,INFNRoma}
\and L.~Gironi\thanksref{MIB,INFNMiB}
\and A.~Giuliani\thanksref{CNRS,DiSAT}
\and P.~Gorla\thanksref{LNGS}
\and C.~Gotti\thanksref{MIB,INFNMiB}
\and G.~Keppel\thanksref{Legnaro}
\and M.~Martinez\thanksref{Roma,INFNRoma,e3} 
\and S.~Morganti\thanksref{INFNRoma}
\and S.~Nagorny\thanksref{LNGS,GSSI,e4} 
\and M.~Nastasi\thanksref{MIB,INFNMiB}
\and S.~Nisi\thanksref{LNGS}
\and C.~Nones\thanksref{CEA}
\and D.~Orlandi\thanksref{LNGS}
\and L.~Pagnanini\thanksref{MIB,INFNMiB}
\and M.~Pallavicini\thanksref{Genova,INFNGenova}
\and V.~Palmieri\thanksref{Legnaro,e5} 
\and L.~Pattavina\thanksref{LNGS,GSSI,e6}} 
\and M.~Pavan\thanksref{MIB,INFNMiB}
\and G.~Pessina\thanksref{INFNMiB}
\and V.~Pettinacci\thanksref{Roma,INFNRoma}
\and S.~Pirro\thanksref{LNGS}
\and S.~Pozzi\thanksref{MIB,INFNMiB}
\and E.~Previtali\thanksref{INFNMiB}
\and A.~Puiu\thanksref{MIB,INFNMiB}
\and C.~Rusconi\thanksref{LNGS,USC} 
\and K.~Sch\"affner\thanksref{GSSI}
\and C.~Tomei\thanksref{INFNRoma}
\and M.~Vignati\thanksref{INFNRoma}
\and A.~Zolotarova\thanksref{CEA} 
}

\institute{INFN - Laboratori Nazionali di Legnaro, Legnaro (Padova) I-35020 - Italy \label{Legnaro}
\and
Materials Science Division, Lawrence Berkeley National Laboratory, Berkeley, CA 94720 - USA\label{LBNL}
\and
Dipartimento di Fisica, Sapienza Universit\`{a} di Roma, Roma I-00185 - Italy \label{Roma}
\and
INFN - Sezione di Roma, Roma I-00185 - Italy\label{INFNRoma}
\and
Dipartimento di Fisica, Universit\`{a} di Milano-Bicocca, Milano I-20126 - Italy\label{MIB}
\and
INFN - Sezione di Milano Bicocca, Milano I-20126 - Italy\label{INFNMiB}
\and
INFN - Laboratori Nazionali del Gran Sasso, Assergi (L'Aquila) I-67010 - Italy\label{LNGS}
\and
Gran Sasso Science Institute, 67100, L'Aquila - Italy\label{GSSI}
\and
Dipartimento di Fisica, Universit\`{a} di Genova, Genova I-16146 - Italy\label{Genova}
\and
INFN - Sezione di Genova, Genova I-16146 - Italy\label{INFNGenova}
\and
CSNSM, Univ. Paris-Sud, CNRS/IN2P3, Universit\'e Paris-Saclay, 91405 Orsay, France\label{CNRS}
\and
DiSAT, Universit\`{a} dell'Insubria, 22100 Como, Italy\label{DiSAT}
\and
IRFU, CEA, Universit\'{e} Paris-Saclay, F-91191 Gif-sur-Yvette, France\label{CEA}
\and
Department of Physics  and Astronomy, University of South Carolina, Columbia, SC 29208 - USA\label{USC}
}

\thankstext{e1}{e-mail: laura.cardani@roma1.infn.it}
\thankstext{e2}{Present address: Max-Planck-Institut f\"ur Physik, 80805, M\"unchen, Germany}
\thankstext{e3}{Present address: Fundacion ARAID and U.  Zaragoza, C/ Pedro Cerbuna 12, 50009 Zaragoza, Spain}
\thankstext{e4}{Present address: Queen's University, Kingston, K7L 3N6, Ontario, Canada}
\thankstext{e5}{Deceased}
\thankstext{e6}{Present address: Physik Department, Technische Universit\"at M\"unchen, D­85748 Garching, Germany}

\date{Received: date / Accepted: date}

\maketitle

\begin{abstract}
The \CupidZ\ experiment searches for double beta decay using cryogenic calorimeters with double (heat and light) read-out. The detector, consisting of 24 ZnSe crystals 95$\%$ enriched in $^{82}$Se and two natural ZnSe crystals,  started data-taking in 2017 at Laboratori Nazionali del Gran Sasso. We present the search for the neutrino-less double beta decay of $^{82}$Se into the 0$_1^+$, 2$_1^+$ and 2$_2^+$ excited states of $^{82}$Kr with an exposure of \exposure\ (\exposureSe). We found no evidence of the decays and set the most stringent limits on the widths of these processes: \LimitFirst, $\Gamma$($^{82}$Se $\rightarrow ^{82}$Kr$_{2_1^+}$)$<6.25 \times10^{-24}$\,yr$^{-1}$,  \LimitSecondTwo\  (90$\%$ credible interval).

\end{abstract}

\section{Introduction}
\label{intro}
The double beta decay is a transition among isobaric isotopes $(A,Z) \rightarrow (A,Z+2) + 2e^- + 2\bar\nu_e$. Despite being among the rarest nuclear processes in Nature, it was observed for eleven nuclei with typical half-lives of 10$^{18}$-10$^{24}$\,years~\cite{BARABASH201552}.
In 1937, Furry hypothesized that double beta decay could occur also without the emission of neutrinos: $(A,Z) \rightarrow (A,Z+2)+ 2e^-$~\cite{Furry}. This process, called neutrino-less double beta decay (\DBD), is forbidden by the Standard Model of Particle Physics as it would violate the difference between the total number of baryons and leptons (B-L)~\cite{Dell'Oro:2016dbc,Feruglio:2002af}. 
Furthermore, \DBD\ is considered a golden channel to probe a fundamental property of neutrinos, i.e. their nature. This transition, indeed, can occur only if (in contrast to all the other known fermions) neutrinos coincide with their own anti-particles, as predicted by Majorana~\cite{PhysRevD.25.2951}. Finally, the measurement of the \DBD\ half-life would help in understanding the absolute mass scale of neutrinos, that today is one of the missing elements in the puzzle of Particle Physics~\cite{Strumia:2005tc}.

\CupidZ\ is the first medium-scale  \DBD\  cryogenic experiment exploiting the dual read-out of heat and light for background suppression~\cite{Azzolini:2018tum}.
The detectors are operated as calorimeters~\cite{Fiorini:1983yj}: each crystal acts as energy absorber converting energy deposits $\Delta E$ into temperature variations $\Delta T$. 
The temperature variation $\Delta T$ is determined by the crystal  thermal capacitance $C$: $\Delta T\propto \Delta E/C$. For a single-particle energy deposition of 1\,MeV, it is possible to observe sizeable signals (hundreds of $\mu$K) only if $C$ is of the order of 10$^{-9}$ --10$^{-10}$\,J/K.
Since in dielectric and diamagnetic crystals $C \propto T^3$ according to the Debye law, such thermal capacitances require the crystals to be cooled down to about 10 mK.
The temperature variations are converted into readable voltage signals using a Neutron Transmutation Doped (NTD) Ge thermistor~\cite{thermistor} glued to the crystal. The resistance of this device shows a strong dependency on the temperature: R(T) $=$ R$_0$\,exp(T$_0$/T)$^{\gamma}$ with $R_0$, $T_0$  and $\gamma$ of about 2\,$\Omega$, 4.2\,K,  and 0.5 respectively. Thus, biasing the thermistor with a small current allows to convert temperature variations in electrical signals with a temperature sensitivity of hundreds of mK per MeV (or hundred of $\mu$V/MeV).

The technological effort of operating tens of massive crystals at cryogenic temperatures is motivated by the advantages that this technique offers in terms of energy resolution, efficiency, and versatility in the choice of the emitter. 
The CUORE experiment~\cite{Artusa:2014lgv,Alduino:2017ehq,Alfonso:2015wka} is successfully operating 988 TeO$_2$ calorimeters for the study of the $^{130}$Te \DBD, proving the feasibility of a tonne-scale experiment based on this technology. According to the CUORE background model, the dominant contribution to the region of interest stems from $\alpha$ particles emitted by the materials in the proximity of the detector~\cite{Alduino:2017qet}. The suppression of the $\alpha$ background is thus the first milestone for next-generation projects aiming at working in an almost background-free environment to increase the discovery potential~\cite{Wang:2015taa,Wang:2015raa,Artusa:2014wnl,Poda:2017jnl}.

The primary goal of the \CupidZ\ experiment is proving that the dual read-out heat/light allows to reject the $\alpha$ interactions, reducing the background in the region of interest for \DBD\ by an order of magnitude.  For this purpose, each calorimeter is coupled to a light detector that enables particle identification exploiting the different light yield of different particles.

The \CupidZ\ detector has been taking data since the end of March 2017 in the underground Laboratori Nazionali del Gran Sasso (LNGS) in Italy. 
The first data release demonstrated the potential of this technology: thanks to the strong background suppression, \CupidZ\ set the most stringent limit on the half-life of the \Se\ decay to the ground state of $^{82}$Kr. Despite the small exposure (1.83\,kg$\cdot$y compared to the 4.90\,kg$\cdot$y collected by NEMO-3~\cite{Barabash2011,Arnold:2018tmo}), \CupidZ\ improved by an order of magnitude the previous limit reaching T$_{1/2}^{0\nu}>$2.4$\times$10$^{24}$\,yr (90$\%$ credible interval)~\cite{Azzolini:2018dyb}.

In this work we search for the \Se\ decay to the 0$_1^+$, 2$_1^+$, 2$_2^+$  excited levels of its daughter nucleus, $^{82}$Kr.
These transitions were already studied using a high purity Germanium detector operated underground at LNGS with an exposure of 3.64$\times$10$^{24}$ emitters$\cdot$yr~\cite{Beeman:2015xjv}. 

Stringent limits on the decay widths were set: \LimitFirstGe, \LimitSecondOneGe, \LimitSecondTwoGe.
The sensitivity of such measurement was mainly limited by the poor detector efficiency (ranging from 0.3 to 3.2$\%$, depending on the chosen signature) and the background level of $9.6 \pm 0.5$\,c/keV/y, ascribed to multi-Compton interactions in the detector.

In contrast to the measurements made with $\gamma$ spectroscopy, in \CupidZ\ we can distinguish the decay with two neutrino emission from neutrino-less double beta decay. In this paper we present the results obtained in the search of the neutrino-less double beta decay with the first data of \CupidZ.


\section{The \CupidZ\ Detector}
\label{sec:detector}
After an extensive R$\&$D on scintillating crystals based on different \DBD\ emitters~\cite{Arnaboldi:2010jx,Beeman:2013vda,Gironi:2010hs,Beeman:2012ci,Beeman:2012jd,Beeman:2011bg,Cardani:2013mja,Armengaud:2015hda,Berge:2014bsa,Cardani:2013dia,Bekker:2014tfa,Artusa:2016maw,Armengaud:2017hit,Buse:2018nzg} the \CupidZ\ collaboration decided to focus on \Se. 
The relatively long half-life of the two-neutrino decay mode (\duenu~\cite{Arnold:2018tmo}) prevents pile-up events in the region of interest despite the slow time response of cryogenic calorimeters ($\sim$ms). The high Q-value of the isotope (\qvalue~\cite{Lincoln:2012fq}) allows to reduce the background due to the environmental radioactivity, that drops above the 2615\,keV line of $^{208}$Tl.
The rather low natural isotopic abundance of \Se\ (8.82$\%$~\cite{SeIsotopicAbundance}) was increased via isotopic enrichment to 96.3$\%$ by the URENCO Stable Isotopes company (Almelo, Netherlands).
The obtained \Se\ was used to synthesize the ZnSe powder, that was later purified and doped using ZnSe(Al) with natural isotopic composition of Zn and Se, in order to enhance the light output. 

The ZnSe powder was used to grow 24 cylindrical ZnSe crystals 95$\%$ enriched in \Se~\cite{Dafinei:2017xpc}. The detector includes also two natural ZnSe crystals, not considered in this work. Since we optimized the crystal shape in order to prevent losses of enriched material, and since we had to reduce the mass of some crystals to discard inclusions and imperfections, the ZnSe crystals feature slightly different size and mass.
The total mass of the 24 \EnrZnSe\ crystals amounts to 9.65\,kg, but two of them are not used for the analysis presented in this paper because of their poor bolometric performance. Thus, the mass considered for the analysis is 8.74\,kg (3.41$\times$10$^{25}$ nuclei of \Se).

The light produced by the ZnSe scintillation (a few $\%$ of the total energy released as heat) escapes the crystal and is recorded using two light detectors.
The fraction of energy converted in form of light depends on the nature of the interacting particle, enabling particle identification and, ultimately, the rejection of the $\alpha$ background~\cite{Pirro:2005ar}. In \CupidZ\ also the light detectors are operated as cryogenic calorimeters, meaning that they convert the impinging photons in temperature variations using NTD Ge thermistors~\cite{Beeman:2013zva}. Nevertheless, in this paper we do not describe the details of the light detectors, as the analysis of coincidences among detectors already provides a sufficient background suppression.

The ZnSe crystals, surrounded by a VIKUITI multi-layer reflecting foil produced by 3M, and interleaved by light detectors, are assembled in five towers using PTFE holders and a mechanical structure made of NOSV copper (produced by Aurubis AG). \\
Each detector was equipped with a Si Joule resistor that periodically injects a reference pulse to correct thermal drifts~\cite{Arnaboldi:2003yp,Andreotti2012}.

More details concerning the detector construction and operation, the  $^{3}$He/$^{4}$He dilution refrigerator, the electronics and data-acquisition can be found in Ref.~\cite{Azzolini:2018tum}.

\section{Expected Signatures in \CupidZ}
\label{sec:signatures}
\begin{figure}[thb]
\begin{centering}
\includegraphics[width=\columnwidth]{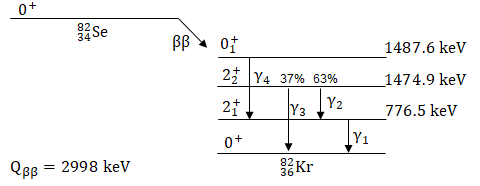}
\caption{Decay scheme of \Se\ to $^{82}$Kr.}
\label{fig:DecayScheme}
\end{centering}
\end{figure}

The decay scheme of \Se\ is shown in Fig.~\ref{fig:DecayScheme}.

If \Se\ decays to the \emph{ground} state of $^{82}$Kr, the two emitted electrons share the entire Q-value of the transition. 
From Monte Carlo simulations, the probability for the two electrons to release the full energy in the crystal, thus producing a peak at the Q-value of the \DBD, is $81.0\pm0.2\%$.

The scenario becomes slightly more complicated if the decay occurs to an excited level of $^{82}$Kr. 
In this case, the energy of the decay is split between the electrons and the $\gamma$ rays emitted during the de-excitation of $^{82}$Kr.
The containment efficiency of the two electrons does not vary significantly, but the probability that a coincident $\gamma$ ray with energy ranging from 698\,keV ($\gamma_2$) to 1475\,keV ($\gamma_3$) releases its full energy in the crystal is very low, leading to an important decrease of the detection efficiency.
The $\gamma$ rays produced in the de-excitation, indeed, can be fully absorbed in the crystal, or escape the crystal and be absorbed in another one, or escape the crystal and scatter in another crystal, or completely escape detection. Depending on the scenario, we expect different signatures. In addition, more decay schemes can result in the same signature, further complicating the analysis.
The redundancy of states, as well as the different detection efficiency of the processes, impose a down-selection of the decay schemes.

First, we exclude from the analysis the events in which a \emph{single} crystal triggers, such as those in which the $\gamma$ rays escape detection. This choice is motivated by the high background produced mainly by the two-neutrino double beta decay of \Se\ (the reader can find a plot of the physics spectrum obtained imposing that a single ZnSe triggered in Ref.~\cite{Azzolini:2018dyb}).

Then, we restrict the analysis to events in which \emph{only two} ZnSe crystals trigger, thus rejecting interactions in three or more crystals. 
The criteria excludes events involving electrons and two $\gamma$'s interacting in three different crystals. This choice is motivated by the very low efficiency of these signatures.

Finally, we discard the remaining signatures with efficiency lower than 0.01$\%$ that would not give a substantial contribution to the analysis. 

To compute the detection efficiency, we simulate 10$^7$ decays in the \CupidZ\ crystals, accounting also for a linear smearing of the energy resolution as a function of the energy (Sec.~\ref{sec:analysis}).  We select events in which two detectors trigger and the energy measured by one of them (E$_{coinc}$) is compatible with the energy of a de-excitation $\gamma$ ray. Then, we search for a peak in the other crystal with energy E$_{main}$ equal to the total energy of the two electrons E$_{\beta\beta}$, or to the sum of the energy of the electrons and another $\gamma$ ray emitted in the same decay E$_{\beta\beta}$+E$_{\gamma_i}$.
The signatures chosen for the analysis presented in this paper and their detection efficiencies are summarized in Table~\ref{Table:signatures}. 
\begin{table}[!htb]
\centering
\begin{tabular}{lccccc}
\hline
\hline
                    &Signature                                                                  &E$_{main}$         &E$_{coinc}$                            &$\epsilon$				&\\
                    &										   &[keV]			&[keV]			      &[$\%$] 				&\\
\hline
\hline
1                  &$\beta\beta_1$ $\mid$ $\gamma_1$                           &2220.5 		&776.5				&1.817$\pm$0.009           &A\\     
\hline
2                  &$\beta\beta_2$ $\mid$ $\gamma_1$                           &1522.1 		&776.5				&0.604$\pm$0.004  		&B\\     
3                  &$\beta\beta_2$ $\mid$ $\gamma_2$                           &1522.1 		&698.4				&0.664$\pm$0.004 	 	&C\\     
4                  &$\beta\beta_2$ $\mid$ $\gamma_3$                           &1522.1 		&1474.9				&0.919$\pm$0.007 		 &D\\     
5                  &$\beta\beta_2$ $\mid$ $\gamma_1$ +  $\gamma_2$ &1522.1 		&1474.9				&0.0141$\pm$0.0004 	 &D\\     
6                  &$\beta\beta_2$ + $\gamma_1$ $\mid$ $\gamma_2$  &2298.6 		&698.4				&0.201$\pm$0.002  		&E\\     
7                  &$\beta\beta_2$ + $\gamma_2$ $\mid$ $\gamma_1$  &2220.5  		&776.5				&0.211$\pm$0.002 		 &A\\     
\hline
8                  &$\beta\beta_3$ $\mid$ $\gamma_2$                           &1509.4 		&776.5				&0.606$\pm$0.006 		&B\\     
9                  &$\beta\beta_3$ $\mid$ $\gamma_4$                           &1509.4 		&711.1			    	&0.660$\pm$0.006  		&F\\     
10                &$\beta\beta_3$ + $\gamma_1$ $\mid$ $\gamma_4$  &2285.9 		&711.1 				&0.196$\pm$0.003 		&G\\     
11                &$\beta\beta_3$ + $\gamma_4$ $\mid$ $\gamma_1$  &2220.5 		&776.5				&0.200$\pm$0.003  		&A\\     
\hline
\hline
\end{tabular}
\caption{Signatures of the \Se\ decays to the excited states of $^{82}$Kr, grouped according to the decay level: $\beta\beta$ is the energy carried away by electrons in the decay to the 2$_1^+$ state ($\beta\beta_1$), to the 2$_2^+$ state  ($\beta\beta_2$), or to the  0$_1^+$  state ($\beta\beta_3$); $\gamma_i$ are the $\gamma$ rays emitted in the de-excitation to the ground state (Fig.~\ref{fig:DecayScheme}); the vertical bar separates the particles releasing their full energy (E$_{main}$) in the 1$^{st}$ crystal, and the particles releasing their full energy (E$_{coinc}$) in the second crystal.The detection efficiency $\epsilon$ is determined by a Monte Carlo simulation. Different decay schemes resulting in the same signature (for example, 1, 7, 11) are labelled with the same letter in the last column; the letter B indicates two states with a slightly different energy $\beta\beta$, that were grouped given the resolution of the detector.}
\label{Table:signatures}
\end{table}

The number of decays N$_i$ corresponding to the i$^{th}$ signature can be written as a function of the exposure $\xi=2.24\times10^{25}$\,emitters$\cdot$yr, the detection efficiency $\epsilon_i$ (Table~\ref{Table:signatures}), the data selection efficiency $\eta$ (Sec.~\ref{sec:analysis}) and the width of the corresponding decay channel ($\Gamma$):
\begin{equation}\begin{split}
N_A &= \eta \xi \cdot \left [\epsilon_1\Gamma_{2_1^+}  + \epsilon_7 \Gamma_{2_2^+}  + \epsilon_{11}\Gamma_{0_1^+}  \right ] \\
N_B &= \eta \xi \cdot \left [\epsilon_2\Gamma_{2_2^+}  + \epsilon_8\Gamma_{0_1^+} \right ] \\
N_C &= \eta \xi \cdot \epsilon_3\Gamma_{2_2^+}  \\
N_D &= \eta \xi \cdot \left [ \epsilon_4\Gamma_{2_2^+} + \epsilon_5\Gamma_{2_2^+}\right ]   \\ 
N_E  &= \eta \xi \cdot \epsilon_{6}\Gamma_{2_2^+}   \\ 
N_F &= \eta \xi \cdot \epsilon_{9}\Gamma_{0_1^+}    \\
N_G &= \eta \xi \cdot \epsilon_{10}\Gamma_{0_1^+} \\
\end{split}\end{equation}

Each signature can be modeled with a function describing the detector response to a monochromatic energy deposit ($\Sigma$) and a flat background component ($\rho^{flat}$). In the case of signatures  B, C, D and F, a gaussian component is added to represent the background from the $^{40}$K line ($\Sigma_{40K}$):
\begin{equation}
\begin{split}
f^A &= N_A  \Sigma_A + (N_A^{bkg,1} \cdot \rho^{flat}_A)\\
f^B &= N_B  \Sigma_B + (N_B^{bkg,1} \cdot \rho^{flat}_B + N_B^{bkg,2} \cdot \Sigma_{40K})\\
\ldots\\
f^F &= N_F  \Sigma_F + (N_F^{bkg,1} \cdot \rho^{flat}_F + N_F^{bkg,2} \cdot \Sigma_{40K})\\
f^G &= N_G  \Sigma_G + (N_G^{bkg,1} \cdot \rho^{flat}_G)\\
\end{split}
\end{equation}

In Sec.~\ref{sec:processing} we describe how data are acquired and processed. In Sec.~\ref{sec:analysis} we derive a model for the detector response $\Sigma$ as a function of the energy and compute the data selection efficiency. Finally, in Sec.~\ref{sec:results} we perform the simultaneous fit of the models $f^A\ldots f^G$ to the data in order to extract the values of $\Gamma_{0_1^+}$,  $\Gamma_{2_1^+}$  and $\Gamma_{2_2^+}$.

\section{Data Collection and Processing}
\label{sec:processing}
The temperature variation produced by 1\,MeV energy deposit in a ZnSe results in a voltage signal of tens of $\mu$V, with typical rise-times of 10\,ms and decay-times ranging from 15 to 60\,ms, depending on the detector.
The voltage signals are amplified and filtered using a Bessel 6 poles anti-aliasing filter with tunable cut-off frequency and gain. More details about the electronics and read-out can be found in Refs.~\cite{Arnaboldi:2018yp,Carniti2016,arnaboldi20018,arnaboldi2015,arnaboldi2010,PIRRO2006672,Arnaboldi:2006mx,Arnaboldi:2004jj,AProgFE}.

The data acquisition system digitizes all the ZnSe channels with a sampling frequency of 1\,kHz and saves the corresponding data on disk in NTuples using the ROOT software framework.
During the measurement we run a software trigger on the acquired data and save the corresponding timestamps in NTuples for the off-line analysis. The trigger algorithm is sensitive to the derivative of the waveforms and its configuration parameters are optimized separately  for each channel~\cite{ThesisDiDomizio,ThesisCopello}.

The analysis presented in this work comprises six \dataset s, each consisting of a collection of physics runs of about two days, plus an initial and final calibration with \THO\ sources to monitor the detector stability (see Table~\ref{tab:livetime}). 
In order to include the Q-value of the \DBD\ in the calibration data, we performed a run with a short living $^{56}$Co source. The source emits $\gamma$ rays up to 3.5\,MeV and has been used at the end of the data taking cycle.
This calibration is used also to study the energy dependency of the energy resolution (Sec.~\ref{sec:analysis}).

The first \dataset\ was devoted to the detector commissioning and optimization, and for this reason it shows the lowest fraction of live-time. This \dataset\ was not used for the \DBD\ analysis presented in Ref.~\cite{Azzolini:2018dyb} because of the poor rejection of the $\alpha$ background due to the variations of the working conditions of the light detectors. 
Concerning the analysis presented in this paper, we do not expect $\alpha$ particles to contribute to the background. Indeed, the searched processes produce events occurring simultaneously in two crystals. Due to the detector shielding, coinciding alpha particle interactions in multiple detectors are very unlikely and are not to be considered a background contribution to the analysis.
Thus, we decided to include also the first \dataset\ to increase the statistics. 

\begin{table}
\centering
\caption{Fraction of time that was spent in physics runs, \THO, $^{56}$Co and Am-Be calibrations (Calib.) and for tests, liquid helium refills of the cryostat, software debug, DAQ problems (Other). In the last column we report the \Se\ exposure (enriched crystals only) collected in each DataSet.}
\label{tab:livetime}       
\begin{tabular}{lcccc}
\hline\noalign{\smallskip}
                            	&Physics      &Calib.  &Other	&Exposure \\
                             	&[$\%$]        &[$\%$]           &[$\%$]  & [emitters$\cdot$yr] \\
\hline
\dataset\ 1      	   	&60.6		& 16.8            &22.6        &3.33$\times$10$^{24}$                 \\  
\hline
\dataset\ 2    	   	&65.0	         &27.6             &7.4         &2.36$\times$10$^{24}$             \\  
\hline
\dataset\ 3 	   	&78.6		&14.1             &7.3          &3.68$\times$10$^{24}$                \\  
\hline
\dataset\ 4           	&83.5		 &14.1             &2.4        &3.19$\times$10$^{24}$                 \\  
\hline
\dataset\ 5           	&82.8		&11.4             &5.8         &4.20$\times$10$^{24}$                 \\  
\hline
\dataset\ 6           	&81.8       		&13.1             &5.1         &5.65$\times$10$^{24}$                   \\  
\noalign{\smallskip}\hline
\end{tabular}
\end{table}

The total ZnSe collected exposure (enriched crystals only) amounts to \exposure, corresponding to 3.05\,kg$\cdot$yr of \Se\ (\exposureSe).
These values account for the dead-time due to detector problems (such as earthquakes or major underground activities) and also for the loss of two enriched crystals due to a non-satisfactory bolometric performance.\\

The collected data are processed off-line using a C++ based analysis framework originally developed by the CUORE-0 collaboration~\cite{Alduino:2016zrl,OuelletThesis,BryantThesis,CushmanThesis}.
The continuous data stream is converted into acquisition windows of 4\,s, with a pre-trigger window of 1\,second to evaluate the detector instantaneous temperature before the pulse occurred.

The data are filtered with a software matched-filter algorithm~\cite{Gatti:1986cw,Radeka:1966} to improve the signal-to-noise ratio. The fluctuations of the pulses amplitude induced by gain instabilities are corrected via reference pulses periodically injected through the Si resistors~\cite{alessandrello1998,alfonso2018}. 
Each peak is modeled by a combination of a Gaussian and an exponential background function. The corrected signal amplitudes are converted into energy with a second order calibration function which was determined using lines between the 511\,keV and 2615\,keV peaks from a  \THO\ source. 

Finally, we compute time coincidences between detectors with a coincidence window of 20\,ms. The time-distribution of real coincidence events was studied using the events collected during the \THO\ calibrations. The 20\,ms window was chosen to ensure a 100$\%$ selection efficiency, at the cost of a possibly larger background due to accidental coincidences. 
Nevertheless, in the next section we show that, given the low detector rate in physics runs, the number of random coincidences is almost negligible.

\section{Data Analysis}
\label{sec:analysis}
To infer the detector response to a monochromatic energy release $\Sigma$, we study the 2615\,keV line produced by the decay of $^{208}$Tl. As explained in Ref.~\cite{Azzolini:2018dyb}, the simplest model describing this peak is the sum of two Gaussian functions $\textit{G}$ with two different $\sigma$ and mean values: 
$\Sigma(\mu_1,\mu_2, \sigma_1, \sigma_2, f_{1,2}) = f_{1,2} \textit{G}(\mu_1, \sigma_1) + (1-f_{1,2})\textit{G}(\mu_2, \sigma_2)$. 
As of today, we do not know the underlying physics behind this bi-Gaussian response. Nevertheless, a multi-Gaussian description of the signal was already observed in other experiments based on cryogenic calorimeters~\cite{Alduino:2016zrl,Europio}.
To account for possible time--variations of the detector response, we fit this model to the 2615\,keV peak in the sum energy spectrum of all the periodical \THO\ calibrations, obtaining $f_{1,2} = 0.83 \pm 0.03$, $\mu_1=2613.88 \pm 0.13$\,keV, $\sigma_1=8.89\pm0.12$\,keV, and $\mu_2=2628.37 \pm 1.42$\,keV, $\sigma_2=15.42\pm0.49$\,keV.

To validate the calibration with \THO\ and characterize the detector response over the range of interest, we perform a run with $^{56}$Co. The $^{56}$Co source emits multiple prominent $\gamma$ rays ranging from 511\,keV to 3451\,keV. This source can not be used as frequently as \THO\ because it needs to be produced by $^{56}$Fe activation via the reaction $^{56}$Fe(p,n)$^{56}$Co. The short $^{56}$Co half-life of about 77\,days results in a usable source live-time of a few weeks. Therefore, frequent calibrations on a monthly basis are perfomed with \THO\ only. $^{56}$Co data are used to validate the calibration function in a wider energy range.

For this purpose, we use the calibration coefficients derived from \THO\ data to calibrate the spectra obtained with $^{56}$Co. 
The most prominent peaks are fit with the bi-Gaussian model. The parameters $f_{1,2}$ and the ratios $\mu_2/\mu_1$ and $\sigma_2/\sigma_1$ were determined using the 2615 keV line only.
This procedure limits the amount of free parameters in the bi-Gaussian model to the mean energy and the energy resolution only. Therefore, $\Sigma(\mu_1,\mu_2, \sigma_1, \sigma_2, f_{1,2})$ can be written as $\Sigma(\mu_1, \sigma_1)$.
In the following we replace the double peak model $\Sigma(\mu_1,\mu_2, \sigma_1, \sigma_2, f_{1,2})$ with the simpler expression $\Sigma(\mu, \sigma)$.

\begin{figure}[!htb]
\begin{centering}
\includegraphics[width=\columnwidth]{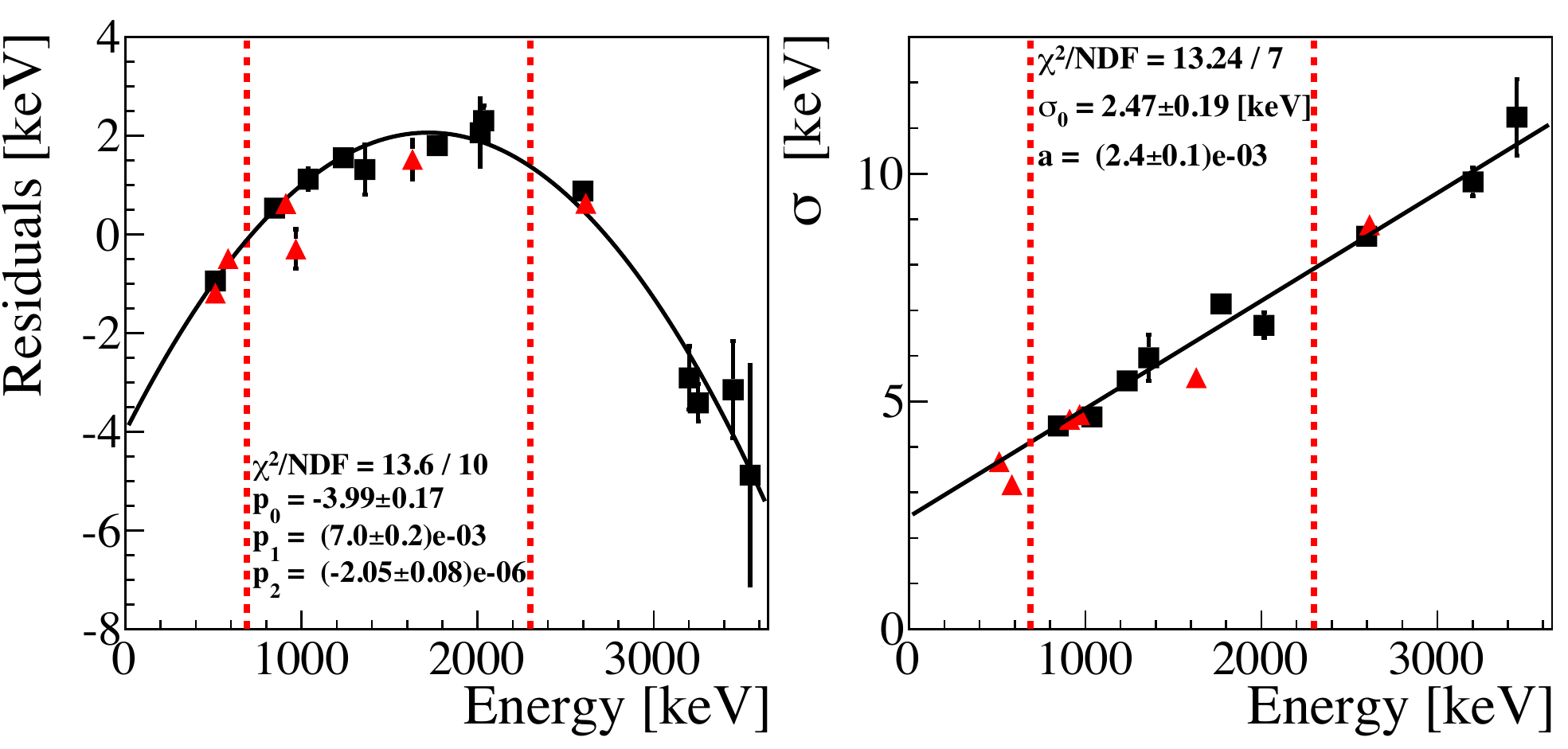}
\caption{Validation of the \THO\ calibration (red triangles) with a $^{56}$Co source (black squares). All the energy spectra of the ZnSe crystals were calibrated using the coefficients derived from the fit of the most prominent \THO\ peaks.
Left: residuals of the calibration, defined as (nominal energy - $\mu$), as a function of the energy; the data obtained with $^{56}$Co are modelled with a parabolic function (black line). Right: the energy resolution measured for different lines is shown as a function of the energy. Data reported in this plot are modeled with a linear function (see text). The vertical dashed lines indicate the region of interest.}
\label{fig:energy_dependency}
\end{centering}
\end{figure}

The parameters $\mu$ and $\sigma$ extracted from the fit of the most prominent $^{56}$Co peaks are reported as a function of the energy in Fig.~\ref{fig:energy_dependency}.
%

This study shows that the residuals obtained using a second order calibration follow a parabolic distribution, meaning that the energy scale could be further optimized. The residuals in the region of interest range from -0.5 to 2.5\,keV. Given the low background rate, we do not expect a relevant impact on the calculated limits resulting from an energy uncertainty of 2.5\,keV. To be more conservative, we check also possible differences in the physics runs with respect to calibration runs by fitting the same model to the most prominent gamma peaks of the physics spectrum. The line with the largest mis-calibration (the 1.46\,MeV line produced by $^{40}$K) features a residual of 2.7\,keV, that will be considered as the uncertainty on the position of the searched peaks (Sec.~\ref{sec:results}).

Fig.~\ref{fig:energy_dependency} shows that the resolution of the peaks scales linearly with energy: $\sigma$(E) $=\sigma_0$ +aE, with $\sigma_0=2.47\pm0.19$\,keV and $a=(2.4\pm0.1)\times10^{-3}$. 
$\sigma_0$ is related to the electronics noise of the detector and is negligible at higher energies, where the loss in energy resolution is dominated by the propagation of phonons in the crystal lattice.
The linear dependency on energy is used to derive the correct signal model at the energy of each signature $\Sigma(\mu, \sigma(E))$.
These models were used to construct the pdf’s and finally fit the selected data. 


The data used for this analysis are selected by imposing a time-coincidence between two crystals in a 20\,ms time-window.
Given an event rate of about 2\,mHz during physics runs, we expect a coincidence rate of 3.2$\times$10$^{-7}$\,events/s in the range from 0 to 10 MeV. 
Taking into account the energy constraints for the studied signatures given in Table~\ref{Table:signatures}, the coincidence rates can be further reduced. With an exposure of  5.74\,kg$\cdot$yr we expect a number of counts ranging from $<$0.08 for signatures A, E, G  to 0.5 counts for signature B.
As a consequence, there is no need to exploit the algorithms for background suppression developed for the search of the \DBD\ to the ground state~\cite{Azzolini:2018yye}.
On the contrary, we apply only a basic cut to the events, selecting windows in which a single pulse is present, to prevent a wrong estimation of the pulse amplitude due to an unpredictable response of the matched filter in presence of multiple pulses. 

The total efficiency comprises the trigger efficiency, the energy reconstruction efficiency, and the efficiency of the quality-cuts applied to the data.
The trigger efficiency, computed on the reference pulses injected by the Si resistor, is defined as the ratio of the triggered to injected pulses. The energy reconstruction efficiency is defined as the number of reference pulses reconstructed within $\pm$3$\sigma$ off the mean energy. Their combined value results in 99.50$\pm$0.02$\%$. 

\begin{figure}[thb]
\begin{centering}
\includegraphics[width=\columnwidth]{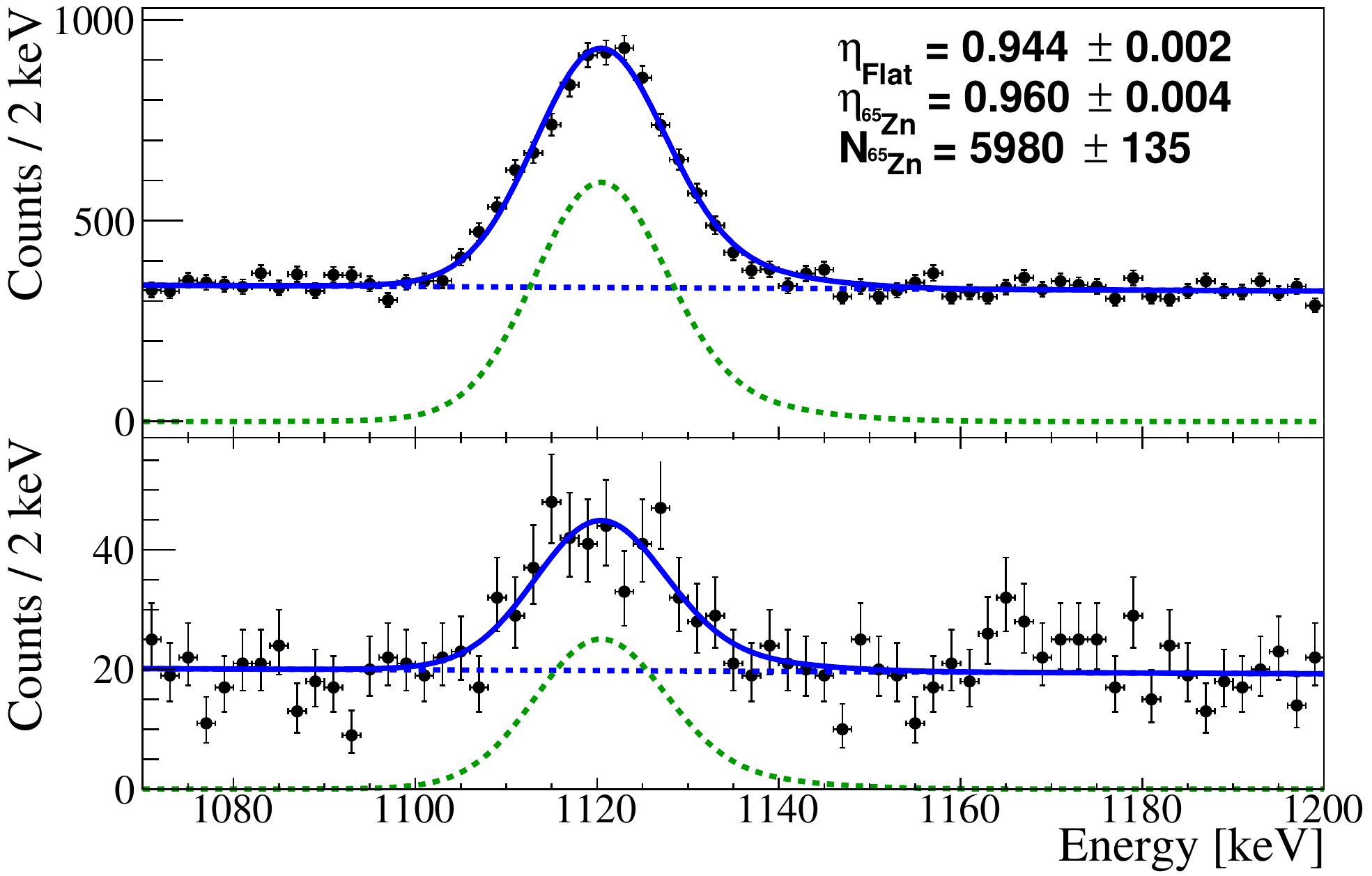}
\caption{$\gamma$ peak produced by the decay of $^{65}$Zn. Top: events in which the acquisition window contains a single pulse. Bottom: events in which the acquisition window contains more than one pulse. The spectra are fitted simultaneously with an unbinned extended maximum likelihood fit (RooFit analysis framework) with two components: the function modeling the detector response $\Sigma(\mu, \sigma(E))$, and an exponential background.}
\label{fig:Efficiency}
\end{centering}
\end{figure}

The data selection efficiency is calculated using the $^{65}$Zn peak which is the most prominent line in the physics-run data with a half-life of 224\,d and a Q-value of 1352\,keV.
The accepted and rejected events are simultaneously fitted with an un-binned extended maximum likelihood fit (with the RooFit analysis framework), resulting in a selection efficiency of 96.0$\pm$0.4$\%$ (Fig.~\ref{fig:Efficiency}). %

The analysis was applied also to the $^{40}$K peak at 1.46\,MeV, resulting in a consistent value for the selection efficiency of 96.0$\pm$1.1$\%$.

The combination of the selection efficiency with the trigger and energy reconstruction efficiencies results in a total efficiency $\eta=95.5\pm0.4\%$.

\section{Results}
\label{sec:results}
The search for the signatures listed in Table~\ref{Table:signatures} starts with data selection as described in Sec.~\ref{sec:analysis}. 
We use a 400 keV analysis window for each signature with the background considered constant in the region of interest.
Furthermore, a 20\,ms  time-coincidence cut is applied to the data in the range of [E$_{coinc}-2\sigma$, E$_{coinc}+2\sigma$] with $\sigma$ given by the fit shown in Fig.~\ref{fig:energy_dependency}.

\begin{figure}[thb]
\begin{centering}
\includegraphics[width=\columnwidth]{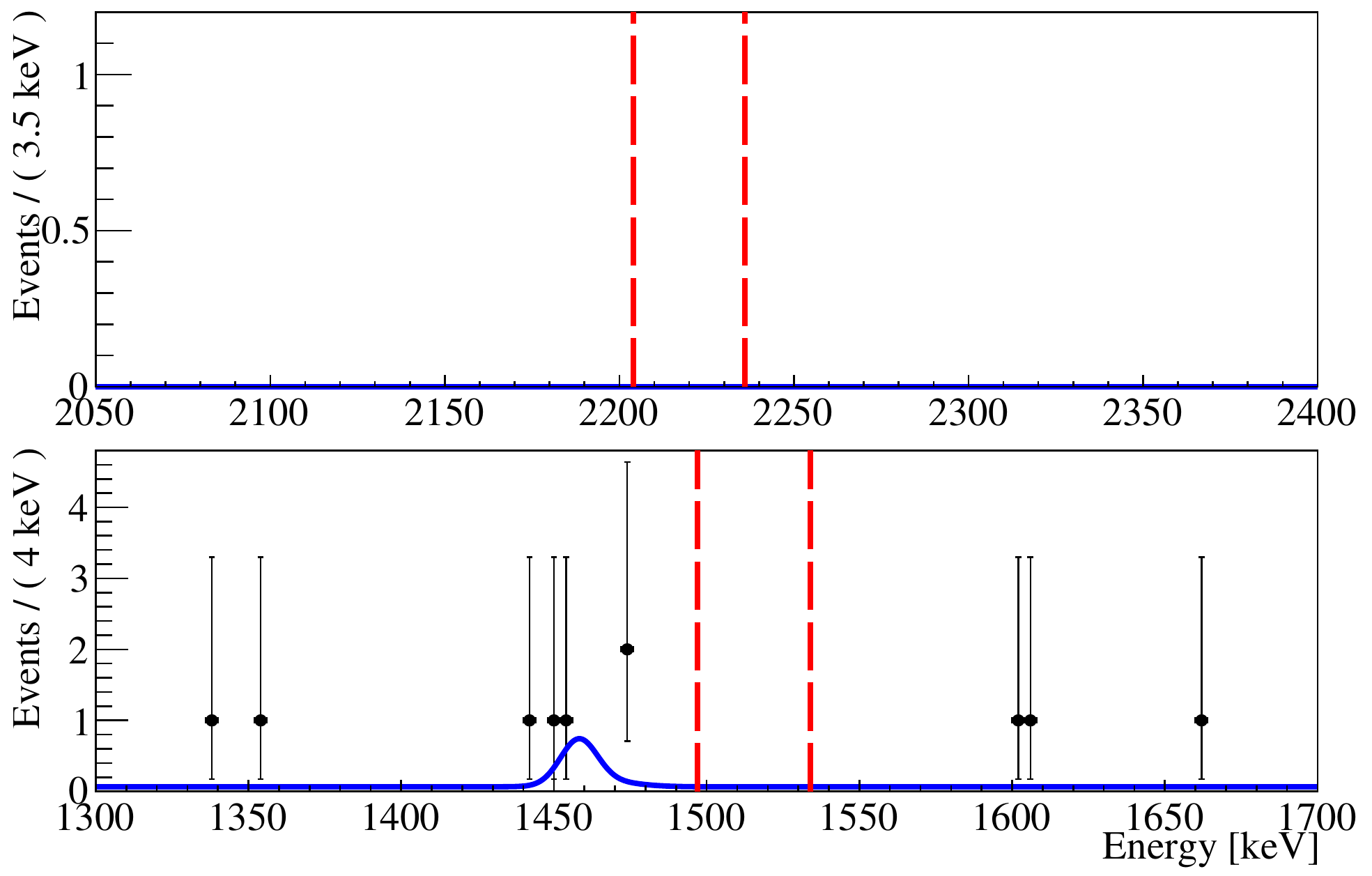}
\caption{Energy spectra of E$_{main}$ corresponding to signatures A (top) and B (bottom) with an exposure of \exposureSe. The red vertical bars indicate a $\pm$2$\sigma$ region centered around E$_{main}$ (Table~\ref{Table:signatures}). The best fit result is shown in blue. In the bottom panel we include also a peaking background to model the excess of events at the energy of $^{40}$K.
}
\label{fig:Spectra}
\end{centering}
\end{figure}

Fig.~\ref{fig:Spectra} shows the resulting spectra for signature A, with less then 0.08 expected counts, and for signature B, with expected 0.5 counts. Other signatures are not shown due to their similarity to signature A.

%

The decay widths are measured using a simultaneous unbinned extended maximum likelihood fit with the following free parameters: $\Gamma_{0_1^+}$,  $\Gamma_{2_1^+}$  and $\Gamma_{2_2^+}$, common to all the signatures,  N$_A^{bkg,1} \ldots$ N$_G^{bkg,1}$ (the number of events ascribed to the flat background in each spectrum), N$_B^{bkg,2}$, N$_C^{bkg,2}$, N$_D^{bkg,2}$ and N$_F^{bkg,2}$(the number of events ascribed to the $^{40}$K peaking background).
The best fit model, shown in Fig.~\ref{fig:Spectra} for signatures A and B, shows no evidence of decays.

The fit accounts also for various sources of systematic errors.
We evaluate the uncertainty related to the calibration function assuming an uniform distribution in the interval [$\mu$-$\Delta$, $\mu$+$\Delta$], with $\Delta=$2.7\,keV, conservatively chosen from the study of the residuals.

The systematic error on $\Gamma_{0_1^+}$,  $\Gamma_{2_1^+}$  and $\Gamma_{2_2^+}$ resulting from $\sigma$ extrapolation in the region of interest, and from the uncertainty on the selection efficiency $\eta$, are estimated by weighting the likelihood by a Gaussian function with the mean fixed to $\sigma$ ($\eta$) and RMS fixed to the uncertainty on $\sigma$ ($\eta$).
We finally integrate the likelihood by a numerical method.

The 90$\%$ credible intervals Bayesian upper limit are set using a uniform prior on the values of $\Gamma_{0_1^+}$,  $\Gamma_{2_1^+}$  and $\Gamma_{2_2^+}$ and marginalizing over the flat and peaking background parameters, obtaining:
\begin{equation}\begin{split}
\Gamma(^{82}Se \rightarrow ^{82}Kr_{0_1^+})&<8.55\times10^{-24}\,yr^{-1};\\
\Gamma(^{82}Se \rightarrow ^{82}Kr_{2_1^+})&<6.25\times10^{-24}\,yr^{-1};\\
\Gamma(^{82}Se \rightarrow ^{82}Kr_{2_2^+})&<8.25\times10^{-24}\,yr^{-1}.
\end{split}
\end{equation}

\section{Conclusions}
In this paper we presented the first background-free search of the neutrino-less double beta decay of \Se\ to the excited states of $^{82}$Kr with an exposure of 2.24$\times$10$^{25}$ emitters$\cdot$yr and we set the most competitive upper limits on the decay widths of the 0$_1^+$, 2${_1^+}$ and 2${_2^+}$ levels. 
The detector is still taking data at LNGS with the aim of reaching a ZnSe exposure of 10\,kg$\cdot$yr, that will allow to further improve this result.

\begin{acknowledgements}
This work was partially supported by the Low-background Underground Cryogenic Installation For Elusive Rates (LUCIFER) experiment, funded by ERC under the European Union's Seventh Framework Programme (FP7/2007-2013)/ERC grant agreement n. 247115, funded within the ASPERA 2nd Common Call for R\&D Activities. We thank M.~Iannone for his help in all the stages of the detector assembly,  A.~Pelosi for constructing the assembly line, M. Guetti for the assistance in the cryogenic operations, R. Gaigher for the mechanics of the calibration system, M. Lindozzi for the cryostat monitoring system, M. Perego for his invaluable help in many tasks, the mechanical workshop of LNGS (E. Tatananni, A. Rotilio, A. Corsi, and B. Romualdi) for the continuous help in the overall set-up design. A.~S.~Zolotorova is supported by the Initiative Doctorale Interdisciplinaire 2015 project funded by the Initiatives d'excellence Paris-Saclay, ANR-11-IDEX-0003-0. We acknowledge the Dark Side Collaboration for the use of the low-radon clean room.
This work makes use of the DIANA data analysis and APOLLO data acquisition software which has been developed by the CUORICINO, CUORE, LUCIFER and CUPID-0 collaborations.
\end{acknowledgements}

\bibliographystyle{spphys}       

\end{document}